\documentclass[useAMS,usenatbib]{mn2e}

\usepackage{times}
\usepackage[dvips]{graphicx}

\title[ Expected masses of merging compact object binaries]
{ Expected masses of   merging compact object binaries 
observed in gravitational waves}

\author[Bulik, Gondek-Rosinska \& Belczynski]
{ T Bulik$^{1}$, D. Gondek-Rosinska$^{2,3,1}$ and K. Belczynski$^{4,5}$\\
$^1$ Nicolaus Copernicus Astronomical Center, Bartycka 18, 00716, Warsaw, Poland\\
$^2$ LUTH, Observatoire de Paris, Place Jules Janssen, 92195 Meudon Cedex, France\\
$^3$ Universite Paris 7 Denis-Diderot, 2 place Jussieu 75251 Paris, France\\
$^4$ Northwestern University, 2145 Sheridan Road, Evanston, IL 60208, USA\\
$^5$ Lindheimer Fellow}

\begin{document}

\date{Accepted  .... Received .....; in original form .....}

\pagerange{\pageref{firstpage}--\pageref{lastpage}} \pubyear{2004}

\maketitle

\label{firstpage}

\begin{abstract}
We use the well tested StarTrack binary population synthesis code to 
examine the properties of the population of compact object binaries.
We calculate the distribution of masses and mass ratios
taking into account weights introduced by observability 
in gravitational waves during inspiral. We find that in the 
observability weighted distribution 
of double neutron star binaries there are two peaks: one for nearly 
equal mass systems, and one for the systems consisting of a low and a high
mass neutron star, $q=0.6-0.7$. The observability weighted distribution 
of black hole neutron star binaries is concentrated on the systems 
with the mass ratio $q=0.3-0.5$, while for the double black hole
binaries the observability weighted distribution is dominated by 
  the massive, nearly equal mass binaries with $q>0.7$.
\end{abstract}

 
\section{Introduction}

We are currently witnessing a large increase in the
sensitivity of the gravitational wave observatories.
LIGO \citep{1992Sci...256..325A} is already taking data, 
the development of VIRGO \citep{1990brada}
shows great advances, GEO600 \citep{1992rgr..conf..184D} and 
TAMA300 \citep{1995gwe..conf..112T} are operational.
In the coming years an even more sensitive Advanced LIGO  
will begin taking data. Out of a number of potential sources 
of gravitational radiation the most promising are
 probably  mergers of  compact object binaries, i.e. binaries
 consisting of black holes (BH) and/or neutron stars (NS).
These are the only sources for which  observations in the electromagnetic domain are
consistent with emission  of gravitational waves.
Present efforts to examine data from gravitational wave detectors
show that such detections rely heavily on availability of accurate
templates. This provides  a case for the
importance of accurate  merger calculations.
The data analysis relies on cross correlating 
the data with a number of templates.  Scanning large volume
of the parameter space requires using a large number of templates
and may hinder detection of a real but low amplitude signal.
Any possibility to limit the amount of templates required 
or to show in which region of the parameter space 
a detection is   most likely may improve chances 
of seeing the gravitational waves.

\begin{table*}
\caption{Population synthesis models. We list the number of 
coalescing compact object binaries produced in each simulation.
For detailed models description see \S\,2.1 and \S\,2.2.}

\begin{center}
\begin{tabular}{lp{7.7cm}r}
\hline \hline
Model & Description &   N produced\\
\hline
A      & standard model described in \S\,2.1 &5761 \\
B1     & zero kicks                  & 21535\\
B7     &  single Maxwellian with $\sigma=50$\,km\,s$^{-1}$  &17747\\
B11    & single Maxwellian with $\sigma=500,$\,km\,s$^{-1}$ & 2155\\
B13    &   \citet{1990ApJ...348..485P}  kicks with $V_k=600$km\,s$^{-1}$ &8270\\
C      & no hyper--critical accretion onto NS/BH in CEs &  4798\\
E1     & CE efficiency: $\alpha_{\rm CE}\times\lambda = 0.1 $  &  894\\
E2     & CE efficiency: $\alpha_{\rm CE}\times\lambda =   0.5 $ & 3489\\
E3     & CE efficiency: $\alpha_{\rm CE}\times\lambda =   2$ &8504\\
F1     & mass fraction accreted in non-cons. MT: f$_{\rm a}=0.1 $ & 2483\\
F2     & mass fraction accreted in non-cons. MT: f$_{\rm a}=  1$ &4644\\
G1     & wind decreased by\ $f_{\rm wind}=0.5 $ & 9395\\
G2    & wind changed by\ $f_{\rm wind}=  2$ & 5517\\
J      & primary mass: $\propto M_1^{-2.35}$ & 8220\\
L1     & angular momentum of material lost in non-cons. MT: $j=0.5 $  & 6660\\
L2     & angular momentum of material lost in non-cons. MT: $j=2.0$  & 5547\\
M1     & initial mass ratio distribution: $\Phi(q) \propto q^{-2.7} $ &  852\\
M2     & initial mass ratio distribution: $\Phi(q) \propto   q^{3}$ & 11225\\
O      & partial fall back for $5.0 < M_{\rm CO} < 14.0 \,M_\odot$ & 4116\\
S      & all systems formed in circular orbits& 4667\\
Z1     & metallicity: $Z=0.01$ & 5199\\
Z2     & metallicity:   $Z=0.0001$ & 7074 \\
\end{tabular}
\end{center}

\end{table*}

Thus  it is important to ask the
following questions: what are the most likely objects
to be observed? what are the most important parameter
sets to explore? In order to answer them one needs to
investigate the properties of population of compact object binaries.
Observations provide us with six double neutron star binaries 
\citep{1999ApJ...512..288T,2003Natur.426..531B}.
The radio selected sample of double neutron star 
binaries is biased toward long lived systems.
However, we do not know any black hole neutron star nor
double black hole binary. Therefore inferring the properties
of the population of compact object binaries solely on the
observations of these few systems may lead to erroneous results.
A different approach - the binary population synthesis
allows to investigate the properties of such systems from a theoretical point of view. Binary
population synthesis requires however, a thorough investigation
of the systematic uncertainties due to parametrization of various
stages of stellar evolution.
The population synthesis studies have already been used
to estimate the rates and properties of the mergers
that can be observed by the gravitational wave observatories
\citep{1997MNRAS.288..245L,1998ApJ...496..333F,1998A&A...332..173P,1998ApJ...506..780B,
1999ApJ...526..152F,1999A&A...346...91B,1999MNRAS.309..629B,2002ApJ...572..407B,
2003Nutz}. It has been shown that
the observed sample will most likely be dominated by the
mergers of double black hole binaries 
\citep{1997NewA....2...43L,2003ApJ...589L..37B}.The distribution of observed 
chirp masses was found to be a very sensitive indicator of the stellar evolution 
model while being relatively not sensitive to the star formation rate history
and cosmological model \citep{2003BBinpress}.
A preliminary study of the distribution of mass
ratios in compact object binaries was presented by \citet{BBK03}.

In this paper we use the StarTrack population synthesis code
to investigate the distribution of masses and mass ratios
in the population of compact object binaries. 
We use a convention where the mass ratio $q$ in a binary system
is defined as the ratio of the lower mass component to the higher mass one and
therefore  is always less than   unity. 
In section 2
we shortly describe the code, and demonstrate the difference between
the volume limited and
flux limited distributions of masses of compact
object binaries. We present the results in section 3, and conclusions in section 
4.

\section{Calculations}

We are using the StarTrack binary evolution code
described in detail by \cite{2002ApJ...572..407B}.
The code is  well tested and has been used
in various astrophysical applications: analysis
of gamma ray-burst progenitors \citep{2002ApJ...571..394B},
tracing of evolutionary history of individual
binaries \citep{2002ApJ...574L.147B}, investigation 
of the mass spectra of compact objects \citep{2002ApJ...567L..63B}.
 {\em StarTrack} population synthesis code was specifically
designed to calculate the merger rates and physical
properties of  compact object binaries. It was compared with
a several other  codes (e.g., 
\citet{1997MNRAS.288..245L}, \citet{1998A&A...332..173P}
  \citet{1998A&A...333..557D} , \citet{1999ApJ...526..152F},
  \citet{2001A&A...365..491N}) . The comparisons showed some differences,
however they were understood within the different model
assumptions. Since {\em StarTrack} was designed to deal mostly
with systems containing NSs and BHs, our input physics was
updated and revised as compared to the other codes in respect
to compact object formation.
As a  result we have recognized new evolutionary NS-NS
formation scenarios, and  we have shown that massive
stellar BH may dominate the population of double compact
objects ($\sim 10 M_\odot$) observed in gravitational waves.
Last, but not least, our predictions of NS-NS Galactic
coalescence rates  \citep{2002ApJ...572..407B} are in  good
agreement with the most recent constraints obtained from
the observed sample of these systems 
\citep{2001ApJ...556..340K,2003ApJ...584..985K}.

\subsection{Standard Model}

Within the code the evolution of single stars is
parametrized by the modified formulae of \cite{2000MNRAS.315..543H}.
The single star evolution includes such stages as
the main sequence, evolution on the Hertzsprung gap,
red giant branch, core helium burning, asymptotic giant branch,
and evolution of helium stars.
Two major modifications include low mass helium star evolution and 
calculation of compact object masses. 
In particular, following a number of studies 
\citep{1981A&A....96..142D,1987A&AS...69..183H,1993RMxAA..25...79A,1995ApJ...448..315W}
we allow the low mass helium stars ($\le 4 M_\odot$) to develop
deep convective envelopes. 
Presence of convective envelope plays an important role in the 
behavior of the donor star in the Roche lobe overflow event, and 
may eventually lead to the development of dynamical instability and 
common envelope (CE) evolution, and possible tightening of the binary orbit.

The original \cite{2000MNRAS.315..543H} formulae are used to calculate 
the final CO core mass of a given compact object progenitor at the time 
of supernova/core collapse event.
We use \citet{1986nce..conf....1W}
 stellar models to obtain mass of the final FeNi 
core corresponding to a given CO core mass.
The FeNi core is collapsed to form proto-neutron star, and then we use 
the results of core collapse hydrodynamical calculation of Fryer (1999) 
to calculate the amount of fall back material and the final mass of the 
newly formed compact object. 
We use the following algorithm to derive the masses of a newly formed 
compact object $M_{\rm rem}$:
\begin{equation}
 M_{\rm rem}=\left\{ \begin{array}{lr} M_{\rm FeNi}& M_{\rm CO} \leq
5\,{\rm M}_\odot\\ M_{\rm FeNi} + f_{\rm fb} (M-M_{\rm FeNi})& 5<M_{\rm
CO}<7.6\\ M& M_{\rm CO} \geq 7.6\,{\rm M}_\odot\\ \end{array}\right.  
\label{masses}
 \end{equation} 
 where 
 $M_{\rm FeNi}$ is the mass of the FeNi core, 
 $M_{\rm CO}$ is the mass of the CO core,
 $M$ is the total mass of the star prior to the explosion, and
$ f_{\rm fb}$ is the fall back factor, $ 0<f_{\rm fb}<1$ 
depending on the mass of the star.
This simple formula represents well the results of detailed 
numerical calculations. We do verify the sensitivity of our results to 
changes in the particular numerical values in equation~\ref{masses}, see
e.g. model O below.  Varying stellar evolution parameters like the strength
of winds, or metallicity leads to different core masses for a star of given 
initial mass and also alters the initial final mass relation for single stars.
We find that NSs are formed without a significant amount of fall back
material, while BHs are formed either directly (prompt collapse of a 
massive star) or through partial fallback of material onto proto-neutron 
star.   

The binary evolution takes into account orbit changes due to
wind mass loss, and tidal interactions.
Wind mass loss rates are adopted from \cite{2000MNRAS.315..543H} and they
depend on the stellar parameters of mass losing component (its composition,
mass and evolutionary stage). Specific mass loss rates are adopted for naked
helium stars, the luminous blue variables and pulsating stars.
In stable mass transfer (MT) calculations we allow for non-conservative evolution. 
We assume that part ($f_a$) of the transferred material to accreted 
onto companion star, while rest is ejected from the system with the 
specific   angular momentum ($j$, expressed in the units of the binary
angular momentum). In our standard model we adopt: $f_a=0.5$ and $j=1$. 
If the RLOF episode is dynamically unstable, we follow the spiral in 
through the common envelope phase. If system avoids the merger, we 
calculate the final orbital separation using standard energy 
conservation based prescription of \citep{1984ApJ...277..355W}. The evolution 
through CE phase depends crucially on the efficiency of the orbital 
energy input into the donor envelope ($\alpha_{\rm ce}$) and the specific 
binding energy of the envelope ($\lambda$). Only the product of these
two largely uncertain quantities enter the calculation, and we use 
$\alpha_{\rm ce} \times \lambda = 1 $ in the standard model, however we also
check the sensitivity of the results to this parameter.
During the CE spiral in we allow for hyper critical accretion onto NSs and 
BHs (e.g., \citet{1986ApJ...308..755B},\citet{1989ApJ...346..847C},
\citet{1993ApJ...411L..33C},\citet{1995ApJ...440..270B}). 
As a result  several tenths of solar mass may be accreted onto 
the compact object, and in particular the top heavy NSs may collapse and 
form BHs.
The full description of the hyper critical accretion treatment is given 
in the appendix of \citet{2002ApJ...572..407B}.
 
Supernovae explosions are treated in detail.
The explosion takes place at the randomly selected place on the orbit. 
We allow for explosions on the eccentric orbits, for uncircularized 
systems. We take into account the instantaneous mass and angular momentum 
loss form the binary system. Also a natal kick is added to the orbital
velocity of the newly born compact object to account for the SN 
asymmetry. Kicks are selected from the bimodal distributions of
 \citet{1998ApJ...505..315C}, a weighted sum of two Maxwellians, one with $\sigma=175$\
km~s$^{-1}$ (80\%) and the second with $\sigma=700$~km~s$^{-1}$ (20\%).
A binary is  either disrupted in the explosions, in which case we stop the
evolution, or if it survives we follow the evolution on the new binary
orbit. 

The initial mass of the primary $M_{zams}^1$ is drawn from a power law
initial mass function (IMF) distribution $\propto M^{-2.7}$ 
\citep{1986FCPh...11....1S} within the range $8$-$100\,M_\odot$.
The secondary mass  is
obtained as $M_{zams}^2 = q M_{zams}^1$, where $q$ is the  mass ratio and is
drawn form a  flat  distribution \citep{1935PASP...47...15K}.
We allow for eccentric initial orbits, and the eccentricities are drawn
from a thermal distribution  $\propto e$ 
\citep{1975MNRAS.173..729H,1991A&A...248..485D}.
Finally, the orbital separation distribution is taken to be flat in $\log a$ 
\citep{1983ARA&A..21..343A}, and separation are chosen from few (so the
stars are not formed at the contact configuration) up to maximum of $10^5$ 
solar radii.  
We evolve our stars for a maximum $T_{Hubble}=15$ Gyrs.
The evolutionary model described above is chosen as our reference (standard) 
model and marked with latter "A" on the following figures and tables.

\subsection{Parameter Study}

In order to asses robustness of the results we investigate
20 extra different models of stellar evolution, where we vary
the parameters describing various stages of stellar and binary evolution.
The  models used are listed in Table 1. The range of models
represents the current state of knowledge and uncertainties 
about the binary evolution.
All models are calculated with $2\times 10^6$ initial binaries each.

In models marked with letter "B" we vary the distribution of natal kicks
compact objects receive when they are formed . This is rather uncertain part of 
evolutionary model as we still do not know the mechanism behind the SN/core
collapse asymmetry \citep{2003PhRvL..90x1101B} .
Therefore, we change the kicks quite drastically, from rather non-realistic
model with no kicks (B1) to the very high kicks of model B11.
The higher the kicks, the less compact object binaries we form, since the 
higher kicks tend to disrupt the progenitor systems.
This is one of the most important parameter as far as the number of compact
object binaries are concerned (close to an order of magnitude change).

Since the CE evolution is another highly uncertain part of our evolutionary
scheme, in models "E" we change the efficiency with which orbital energy is
transformed into the unbinding the envelope of the donor star, while in
model "C" we turn off the accretion onto compact objects during that very 
short lived phase.
In models with small CE efficiency (E1-2) it is found that the number of
compact object binaries are significantly reduced. This is due to the fact,
that many binaries, evolving through the CE phase, will merge, thus aborting
compact object binary formation. 
On the other hand, increase of the efficiency (E3) or shutting down the
accretion at CE phase do not play a very important role on the number of
formed compact object binaries.

In models "F" and "L" we consider the results for different treatment of the
stable MT phases. In particular, in model F2 we consider the case of
conservative evolution (all mass and angular momentum transfered to the
companion). Change of the MT mode from the non-conservative (standard model) 
to conservative evolution (F2), does not change the numbers by much. 
The model F1 with highly non-conservative evolution, decrease the numbers of
formed compact object binaries rather significantly, but it is rather
improbable, since the estimated material loss is probably not as high as
assumed in model F1 \citep{1989A&A...226...88M}.
Since we have adopted quite a large value for specific angular momentum of
lost material in non-conservative MT episodes, we should really concern only
the model L1 with lowered specific momentum leaving the binary. As we see
from Table 1, the numbers in model L1 are almost the same as for standard
model. We thus conclude, that the treatment of non-conservative MT phases  
does not have a great influence on the number of compact object binaries.

Winds of massive stars may play an important role on the population of
compact objects. 
In model G1 we decrease all the wind mass loss rates by factor of 2. 
The smaller the winds, the more massive compact objects formed, and more BHs
are formed as compared to NSs.
However, the total number of compact object binaries is basically 
unchanged in this model.
 
The flatter IMF slope of model J increases slightly (as slope was not
changed by much) the number of heavy stars (progenitors of compact objects)
and thus leads to a slight increase in the number of compact object
binaries. The IMF slope for massive stars is rather well determined 
\citep{2003ApJ...598.1076K}, and
as expected the small change does not affect the population.
 
Change of initial mass ratio distribution may have severe effect on the
numbers of compact object binaries (models M1-2). In model M1 most of the
progenitor systems are formed with extremely small mass ratios. Therefore,
once they reach first MT phase, it is usually dynamically unstable, leading
to spiral in and merger of components, aborting the formation of compact
object binary. This explains the small number of formed binaries in model M1,
and warrants survival of the systems (large numbers of compact object
binaries) with rather equal masses in model M2. 
Since the initial mass ratio distribution is not easily measurable and
constrained, models M1-2, although rather extreme, should be taken account
in further analysis. 

In the end, several other models, do not have much influence on the 
production efficiency of compact object binaries. These include models 
with different metallicities (Z1-2), different assumption on initial 
eccentricities (S) and finally the model in which we change the regime 
of the fall back in formation of compact object binaries (O). 

The masses of the compact objects are strongly affected in some of the models.
In particular the masses change quite drastically in models G1 and G2 where the stellar winds
are changed. Decreasing the stellar winds (model G1) allows the star to develop 
more massive  cores and consequently leads to higher
masses of the compact objects formed.  
A similar effect is connected with decreasing the metallicity (models Z1 and Z2),
since lower metallicity stars have smaller winds.
On the other hand within the model G2, 
where the winds are artificially increased, massive stars do not have the time
 to develop massive cores and no compact objects above $3\,M_\odot$ are produced.   
The masses of the compact objects  are also affected by varying parameters 
in equation \ref{masses}. In model O we increase  the upper bound of the fall-back range 
to $14\,M_\odot$. this leads to smaller masses of the black holes produced as even for high
mass cores some fraction of the mass is still expelled. 
The population of compact object binaries in model E1, with reduced 
common envelope efficiency contains additional systems with  
massive black holes.  
Within model C  we turn off the hypercritical accretion 
onto compact objects in common envelope events. This primarily influences the masses
of neutron stars and low mass black holes as they have no possibility to 
increase significantly.

\subsection{Distributions of masses and mass ratios}

In the output we note the masses of the compact objects in each binary
and the lifetimes: the stellar lifetime 
from the formation at the zero age main sequence to formation of
a double compact object, and the lifetime as a double compact object binary 
until it   merges due to gravitational wave emission. We denote the sum of the two 
lifetimes as the total lifetime of the binary $T$.

In this calculation we assume for simplicity that the space is Euclidean.
We denote the masses of the components in each binary 
 $m_1^i$ and $m_2^i$, the mass ratio is $q= m_1^i/m_2^i<1$.
The formation rate of compact object binaries with a given mass ratio $q$,
the mass of the primary $m_2$ (the greater of the two masses), and the lifetime 
$T$  at a given cosmic time $t$ is  
\begin{eqnarray}
{ dF(m_2,q,T,t)\over dm_2 dq dT  } &=&{S(t) f_{sim}\over \left<M_*\right> 
N_{tot} }   \times \label{prod} \\
  & \times&  \sum_{i=1}^{N_{CCOB}} \delta(m_2-m_2^i) \delta(q-q^i) \delta(T-T^i) \nonumber \, ,
\end{eqnarray}
where $S(t)$ is the star formation rate at the time $t$, 
$f_{sim}$ is the fraction 
of stars out of a total population that we simulate,  $\left<M_*\right>$ is the 
average mass of a binary in the stellar population, and 
$N_{CCOB}$ is the number of coalescing compact 
object binaries formed in a simulation of $N_{tot}$ binaries.
Our aim is to calculate the observed merger rate by an observer on Earth 
at present, which we denote as $t_0$. 
The coalescence rate  
at a distance $r$ from the observer is then given by
\begin{equation}
{df_{coal}(r)  \over dm_2 dq } = \int dt'  {dF(m_2,q,t',t_o-r/c-t')\over dm_2 dq dT} 
\label{coal}\, .
\end{equation}
Inserting equation \ref{coal} into \ref{prod} we obtain
\begin{eqnarray}
 {df_{coal}(r)  \over dm_2 dq }& =&
 {f_{sim}\over \left<M_*\right>} N_{tot}^{-1}  \times \label{coall} \\
&\times & \sum_{i=1}^{N_{CCOB}} \delta(m_2-m_2^i) \delta(q-q^i)  S(t_0-r/c-T^i)  \, . \nonumber
\end{eqnarray}
The observed rate is obtained by integrating equation \ref{coall} over the 
volume in which the binaries are observable 
\begin{equation}
{dR\over dm_2 dq} = \int_{V(m_2,q)} dV {df_{coal}(r)  \over dm_2 dq } 
\label{rate}
\end{equation}
We note that for a constant  star formation rate   the 
lifetimes of the binaries do not enter the observed rate.

We first calculate a volume limited  distribution of masses, i.e.
we assume that all binaries coalescing in  a given volume $V$ are observable 
regardless of the mass $m_2$ and the mass ratio $q$.
This corresponds to observing for example 
the entire population of a given galaxy or a galaxy cluster.
Here we also  assume that the star formation history was constant.
The normalized volume limited  distribution of masses and mass ratios is 
\begin{equation}
P(q,m_2) = N_{CCOB}^{-1} \sum_{i=1}^{N_{CCOB}} \delta(q-q^i)\delta(m_2-m_2^i)\, ,
\label{vol}\, .
\end{equation}

In the case of realistic detectors the volume of integration
in equation \ref{rate} will depend on $m_2$ and $q$.
Our calculation of  the distribution of masses and mass ratio  relevant for
detecting merging binaries with   gravitational waves follows
the calculations presented earlier in \citep{2003ApJ...589L..37B}.
Here again we assume that the star formation rate was flat and
that the Universe is Euclidean and uniformly filled
with stars. 
The signal to noise in high frequency gravitational wave detectors from an
inspiral of a stellar mass binary is
given by \citep{1993PhRvD..47.2198F,1994Bon,1998PhRvD..57.4535F}
 \begin{equation}
(S/N) = {A_i\over d } \left[{ {\cal M}\over M_\odot}\right]^{5/6}\, ,
\label{sn}
\end{equation}
 where
${\cal M} = (m_1 m_2) ^{0.6} (m_1+m_2)^{-0.2}$ is the chirp mass,
$d$ is the  distance, and the $A_i$ 
depends on the details of a particular detector.  
Thus a coalescence of a 
binary with a chirp mass ${\cal M}^i$ will be visible
 up to a distance proportional to $({\cal M}^i)^{5/6}$
and the volume of integration in equation \ref{rate} 
  will be    $V^i\propto ({\cal M}^i)^{5/2}$. 
The observability weighted distribution of masses and mass ratios 
 is therefore
\begin{equation}
P_{obs}(q,m_2) = K^{-1} \sum V_i\delta(q-q^i)\delta(m_2-m_2^i)\, ,
\label{obsd}
\end{equation}
where $K=\sum_i V_i$. Again the lifetimes of the binaries do not
enter the weights in equation~\ref{obsd} because of assumption of constant 
star formation rate history.  This distribution is more realistic 
as it corresponds to a case of an instrument with 
the sensitivity allowing it to detect binary coalescences in a large ensemble of galaxies.
Relaxation of the assumption  of constant star formation and taking 
into account a realistic cosmological model  has 
been discussed in \citep{2003BBinpress} for the case of the 
distribution of observed chirp masses and was shown not to be significant.

\begin{figure*}
\begin{center}
 
\includegraphics[width=0.85\columnwidth]{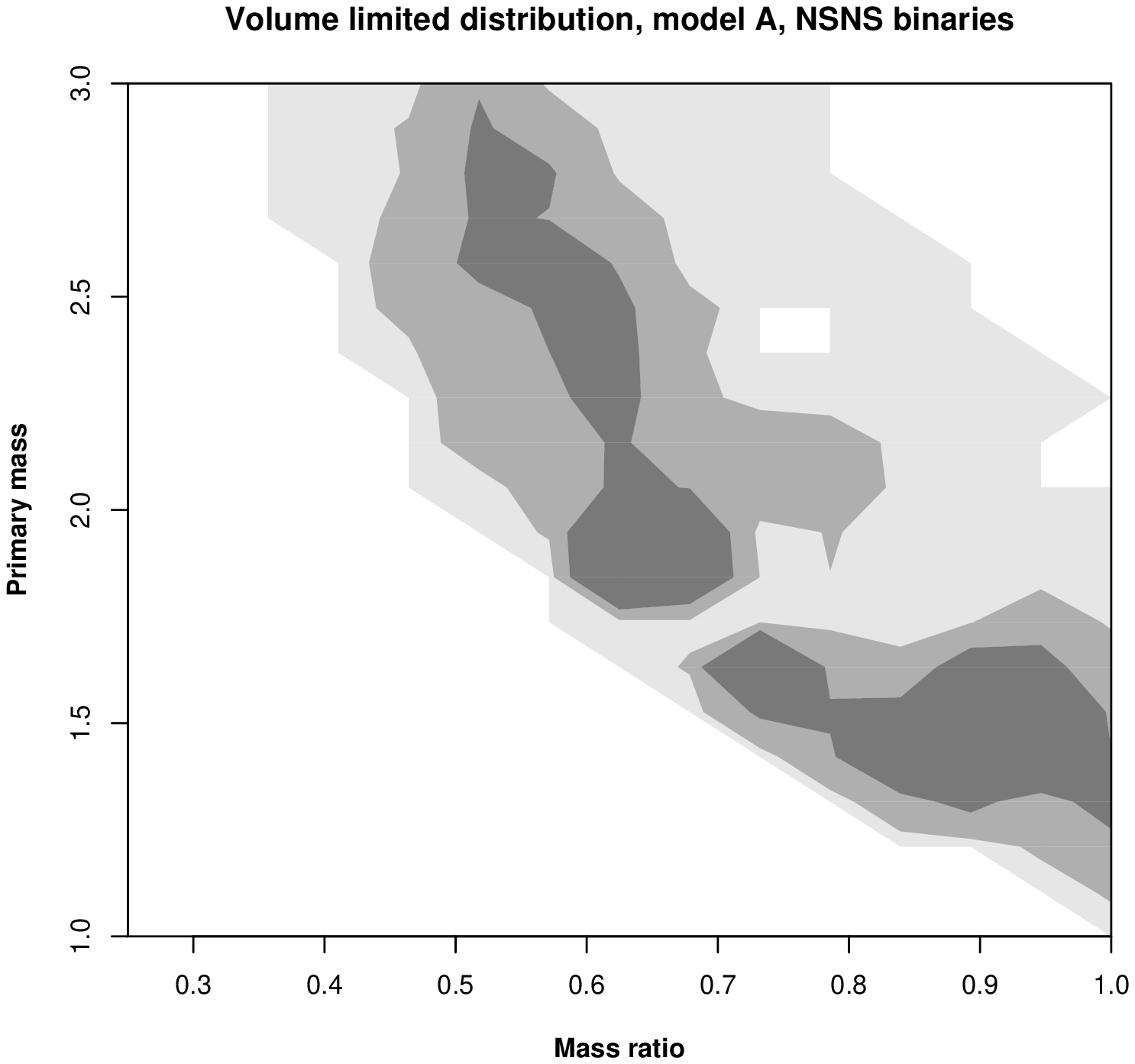}
\includegraphics[width=0.85\columnwidth]{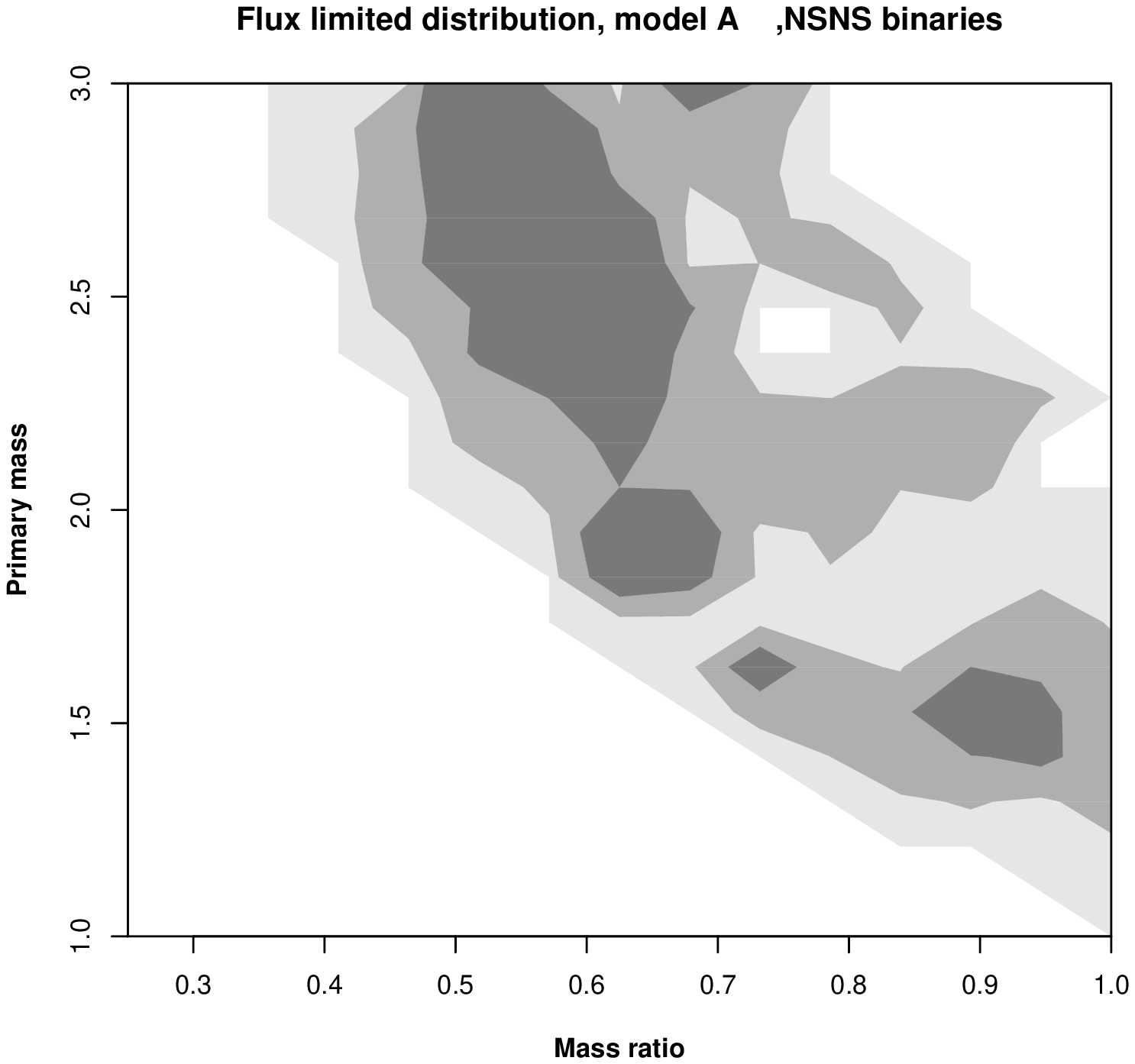}\\
\includegraphics[width=0.85\columnwidth]{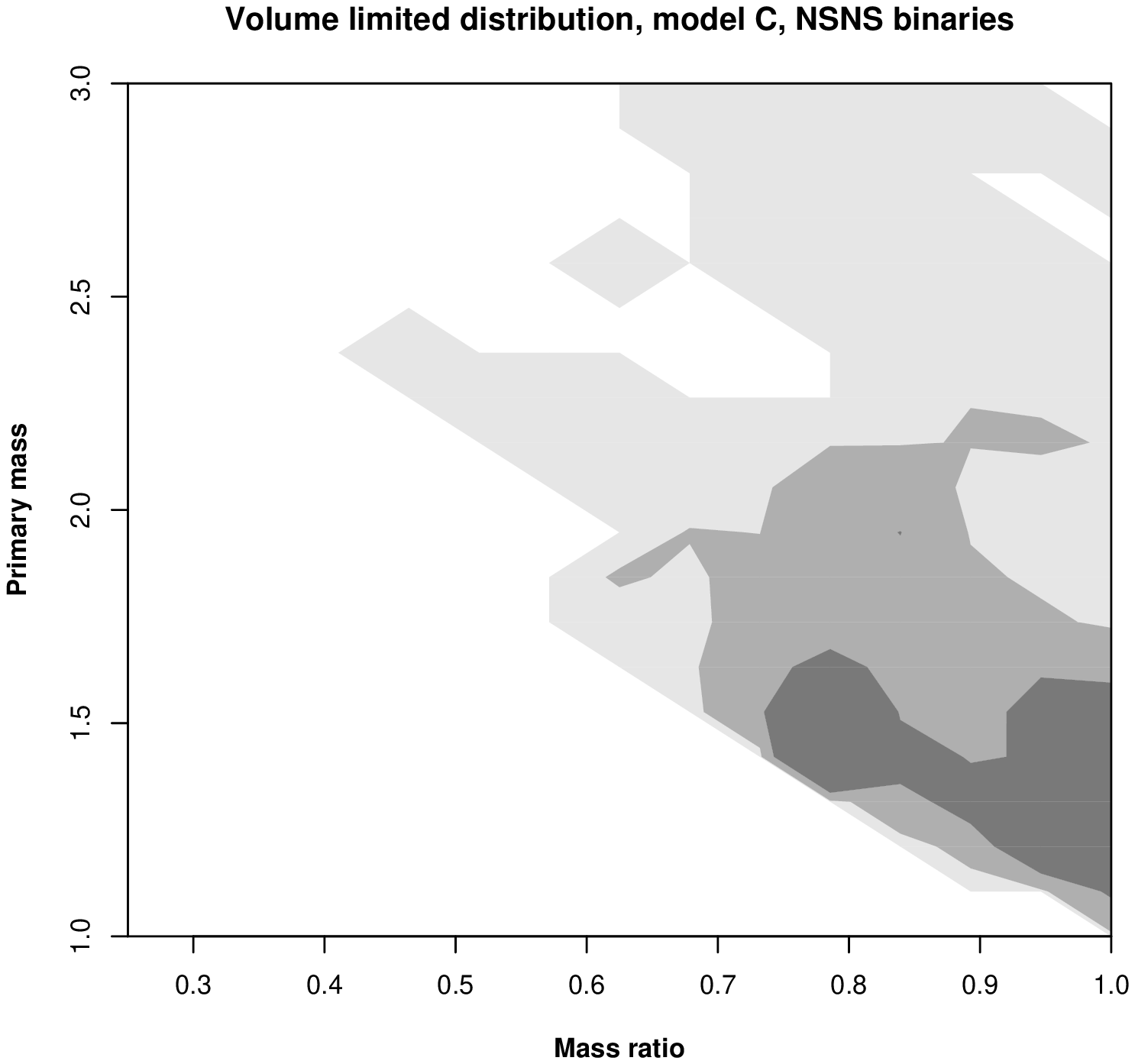}
\includegraphics[width=0.85\columnwidth]{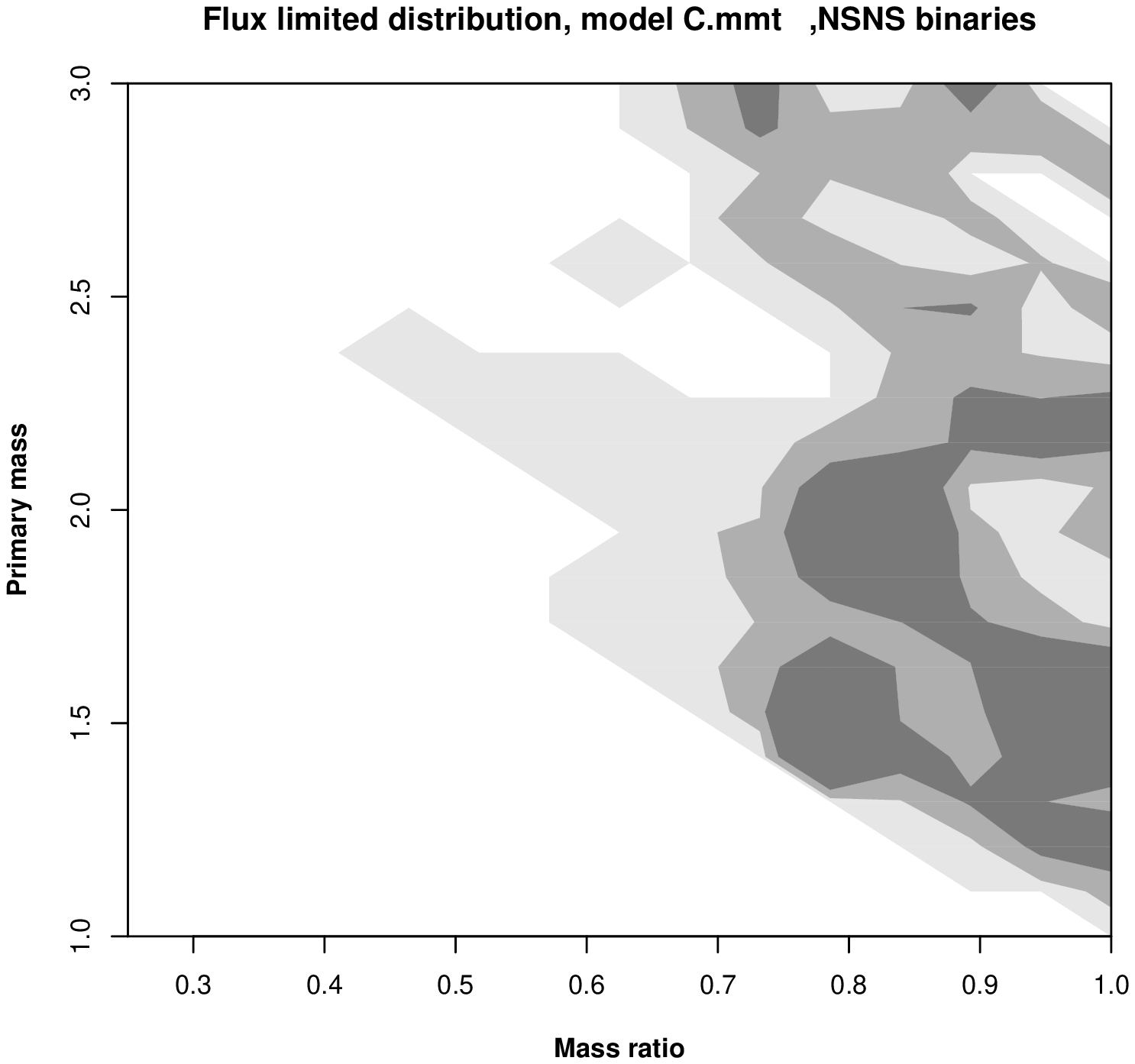}\\
 
\end{center}
\caption{The volume limited  and observability weighted distributions of 
parameters of compact NSNS binaries obtained within model A - top panel 
and model C - bottom panel. The  region in dark gray encompasses
68\% of the systems, the medium gray corressponds to 95\%,
and the light gray corresponds to all binaries in the simulation.}
\label{nsns}
\end{figure*}

\begin{figure*}
\begin{center}
\includegraphics[width=0.85\columnwidth]{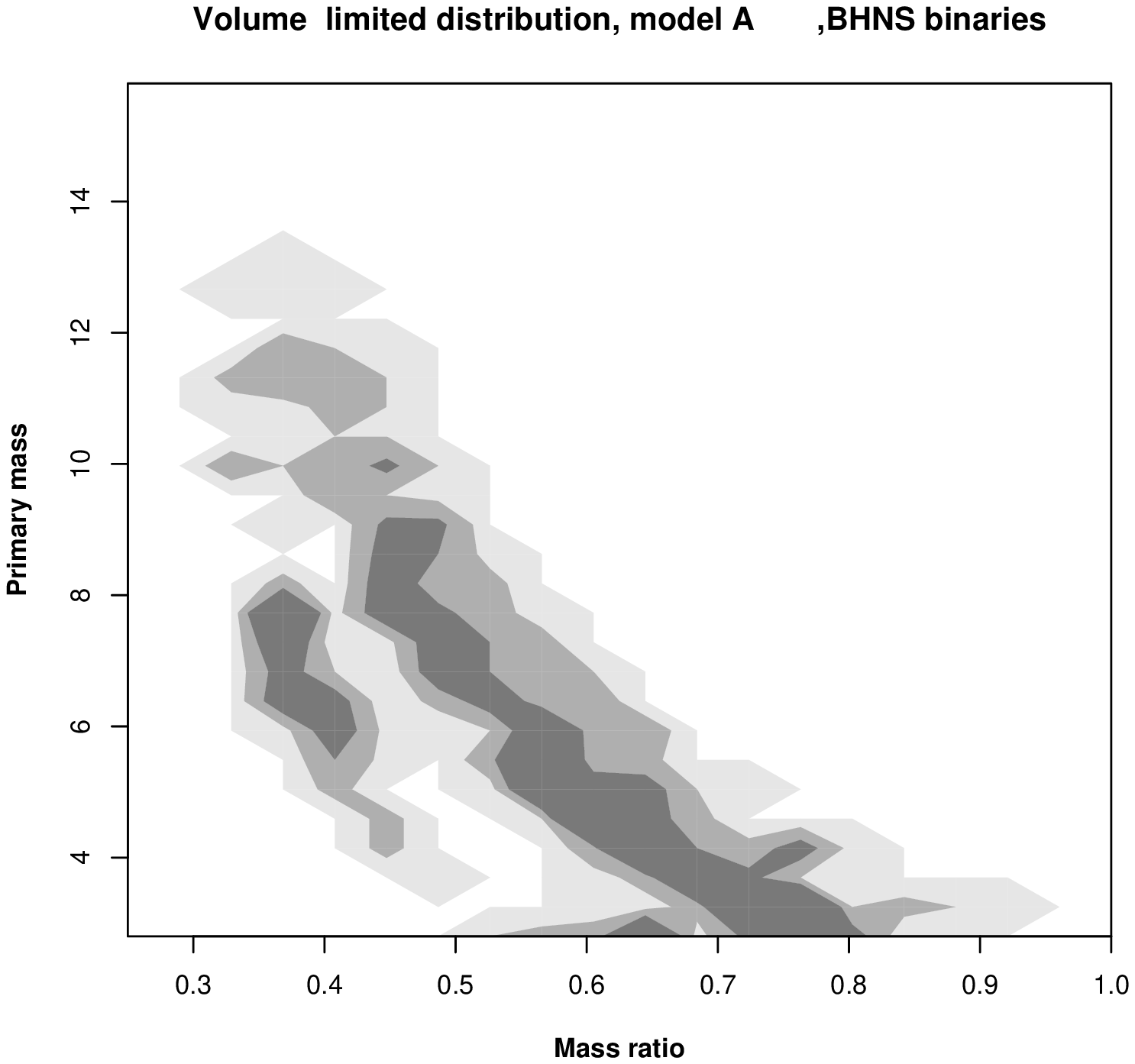}
\includegraphics[width=0.85\columnwidth]{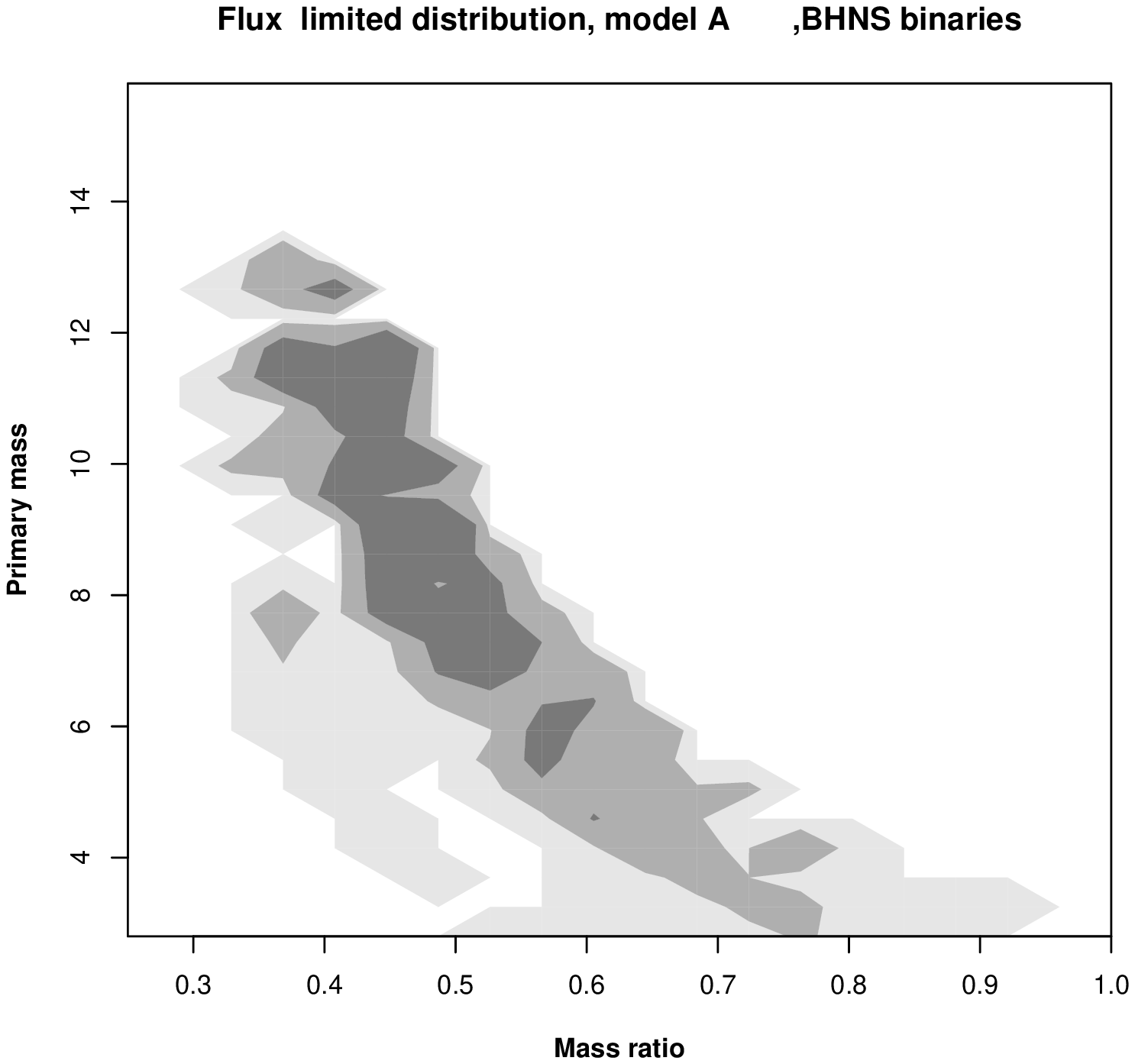}\\
\includegraphics[width=0.85\columnwidth]{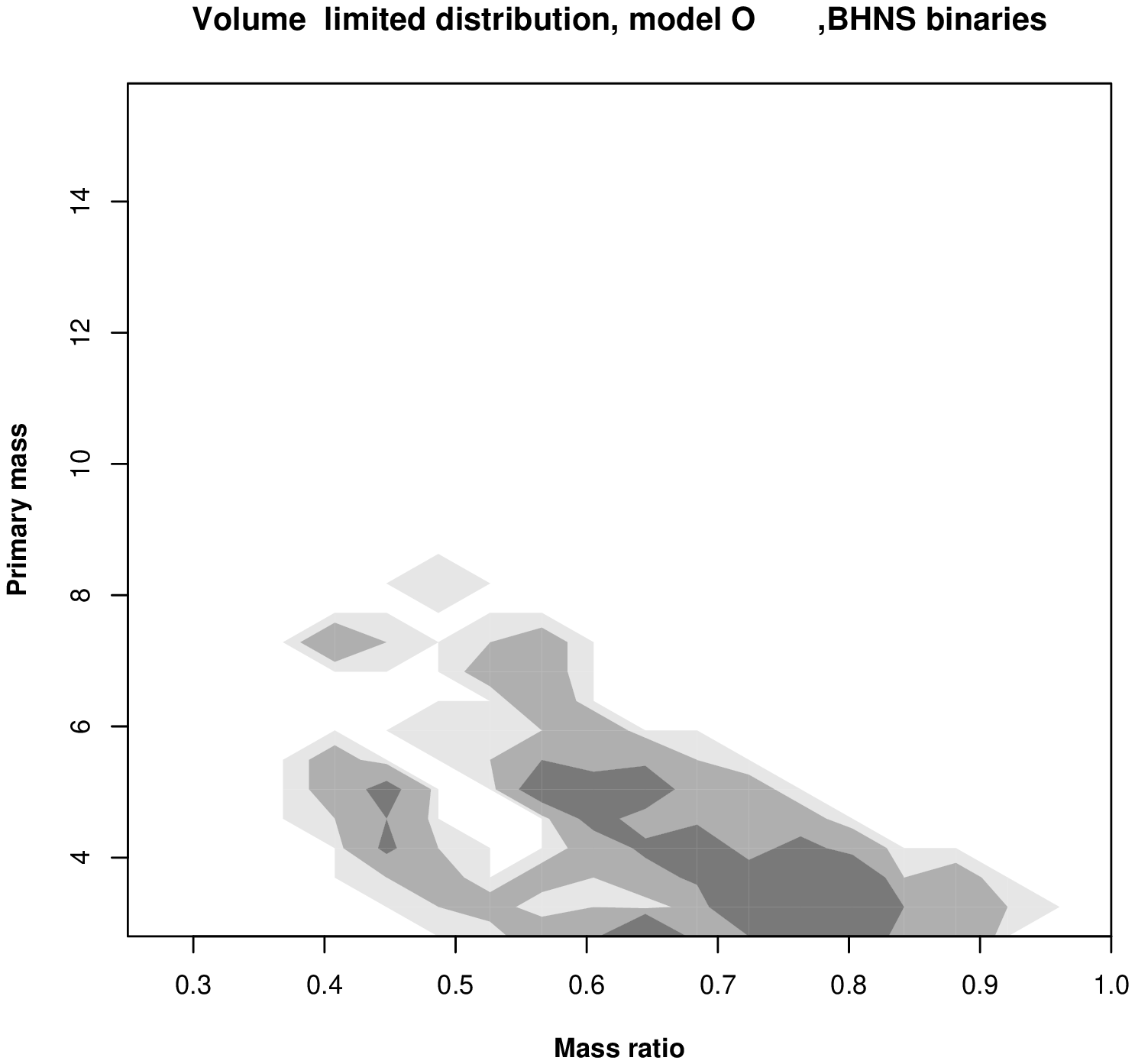}
\includegraphics[width=0.85\columnwidth]{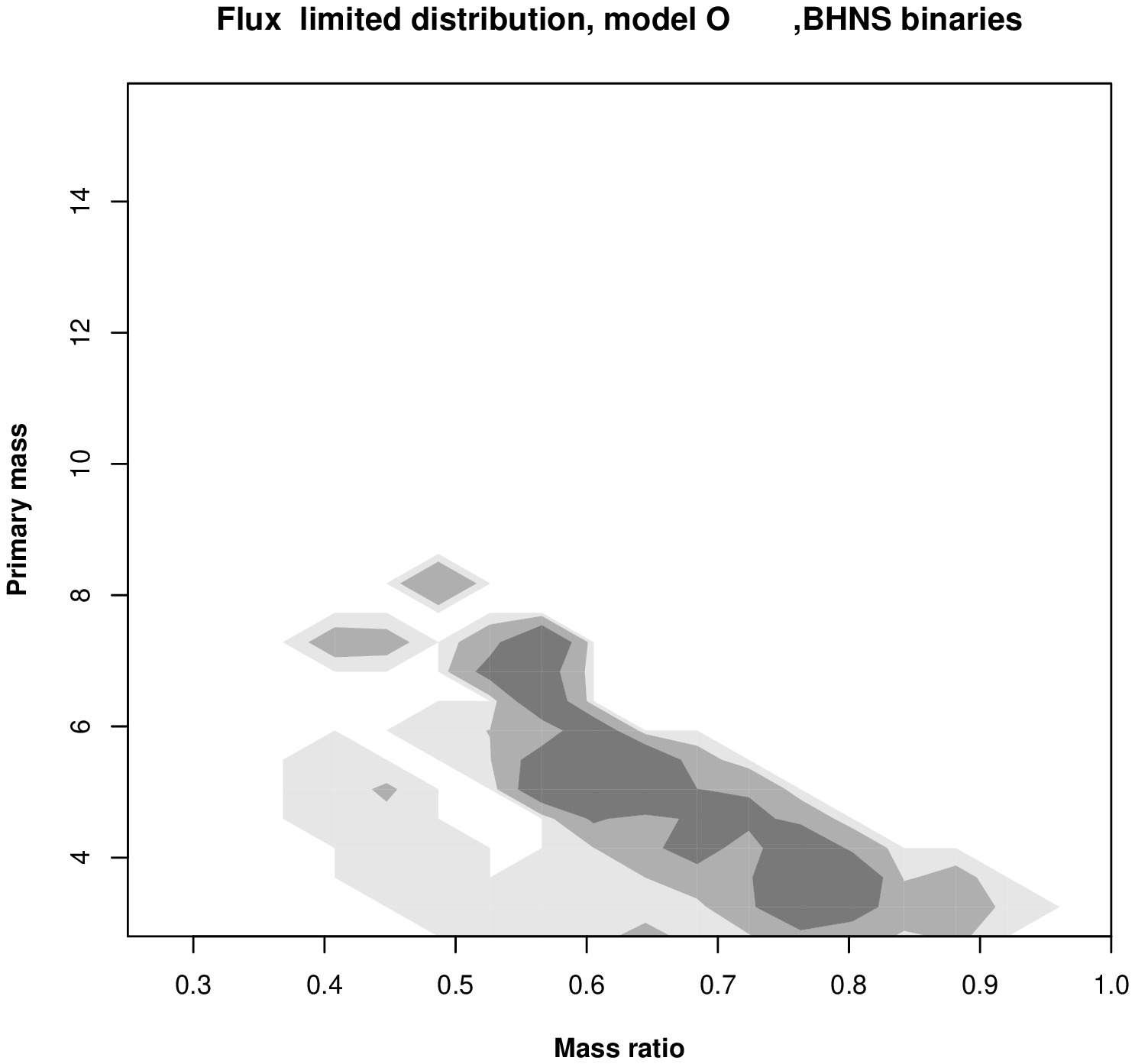}\\
\includegraphics[width=0.85\columnwidth]{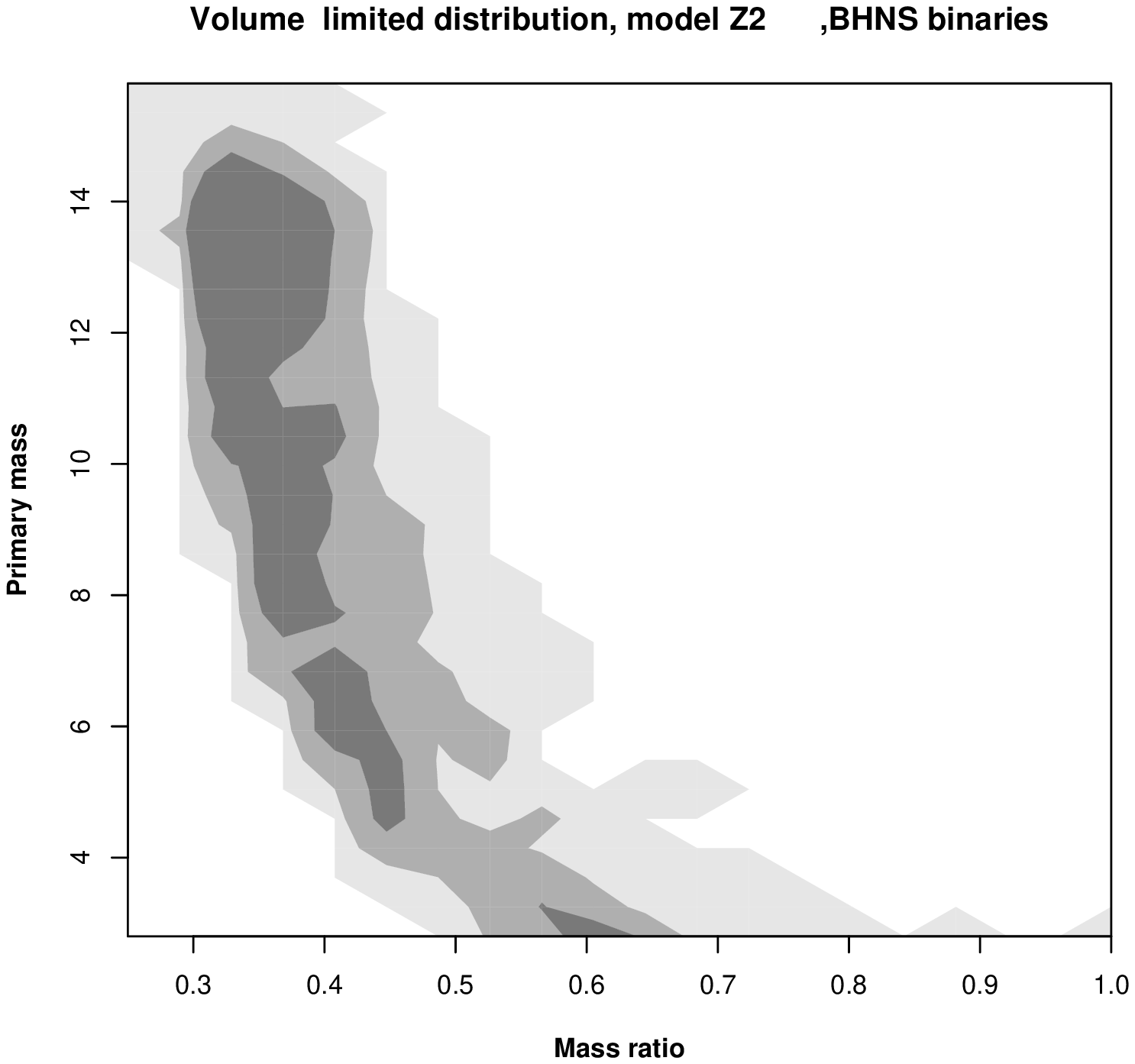}
\includegraphics[width=0.85\columnwidth]{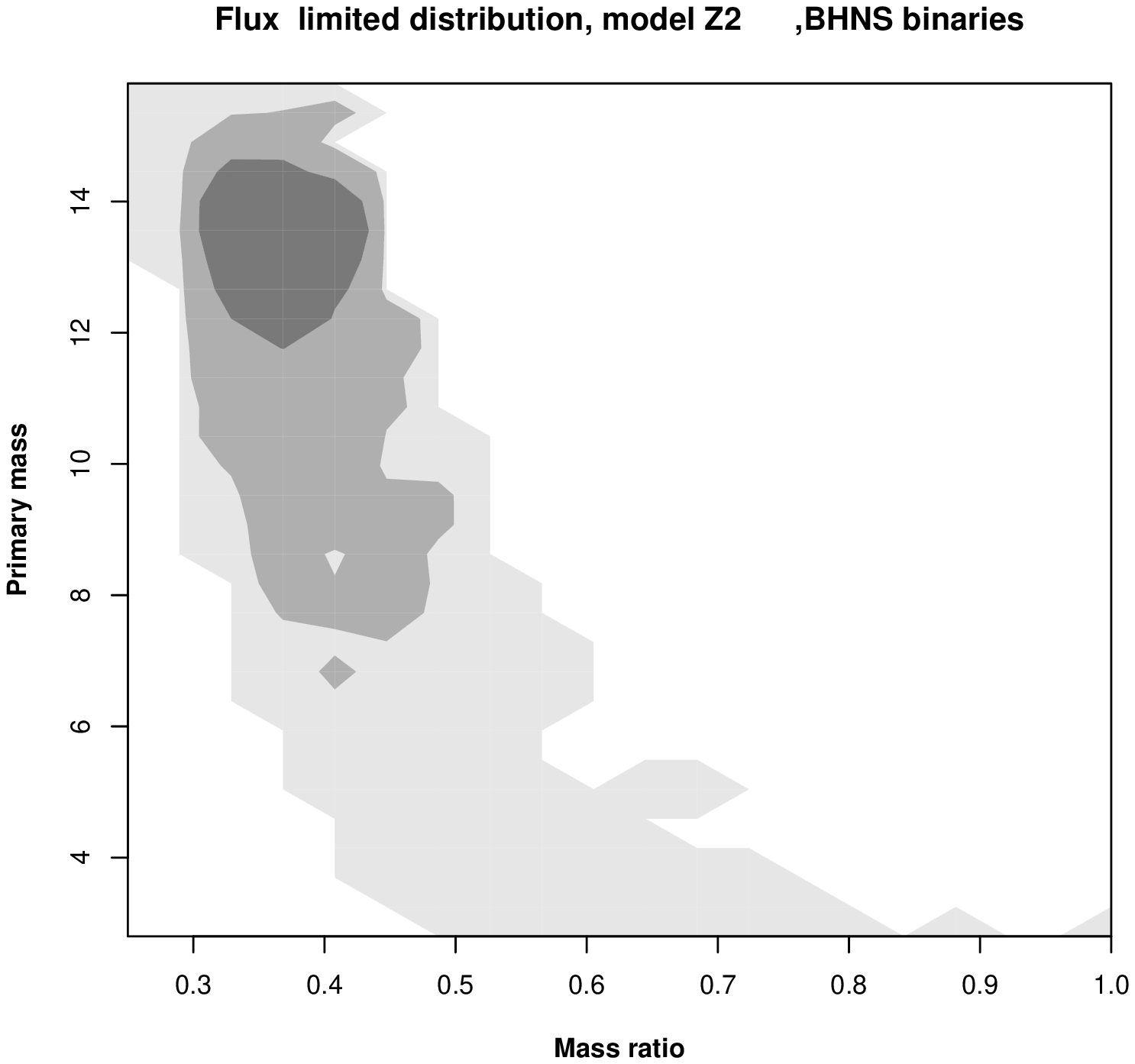}\\
\end{center}
\caption{The volume limited  and observability weighted distributions of 
of the parameters of BHNS binaries within model A (top panel),
model O (middle panel) and model Z2 (bottom panel). The  region in dark gray encompasses
68\% of the systems, the medium gray corressponds to 95\%,
and the light gray corresponds to all binaries in the simulation. }
\label{bhns}
\end{figure*}

\begin{figure*}
\begin{center}
\includegraphics[width=0.85\columnwidth]{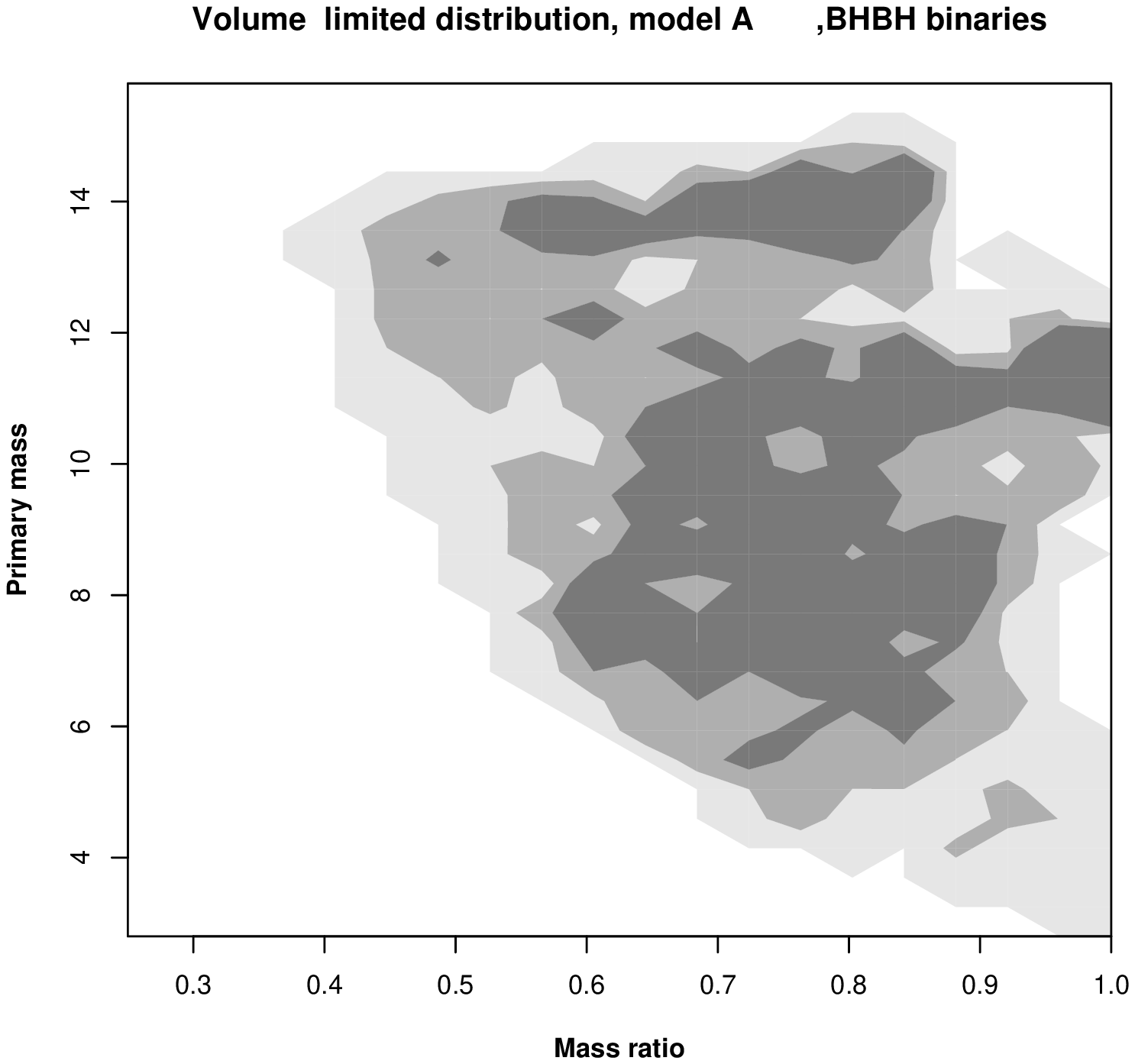}
\includegraphics[width=0.85\columnwidth]{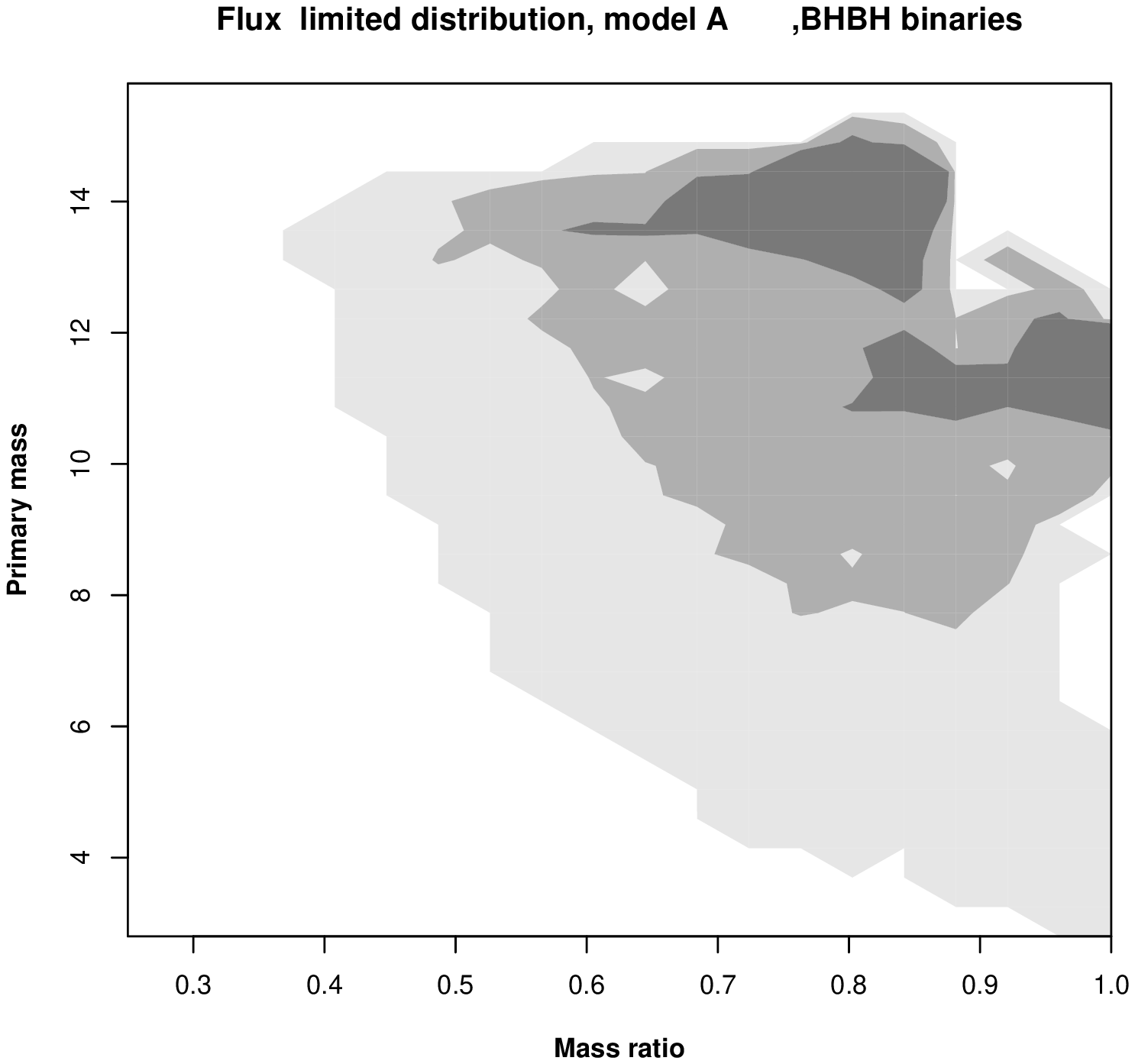}\\
\includegraphics[width=0.85\columnwidth]{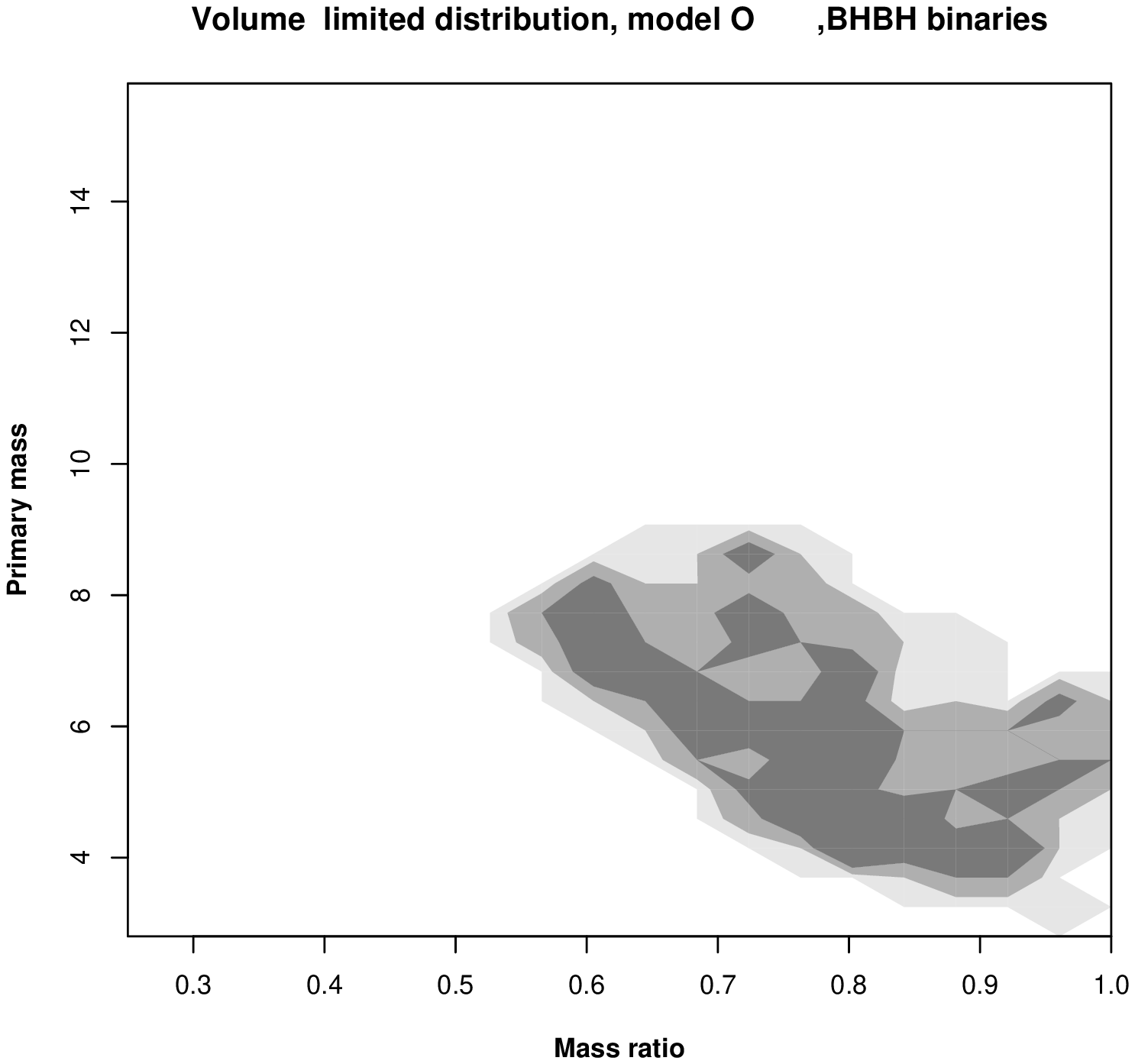}
\includegraphics[width=0.85\columnwidth]{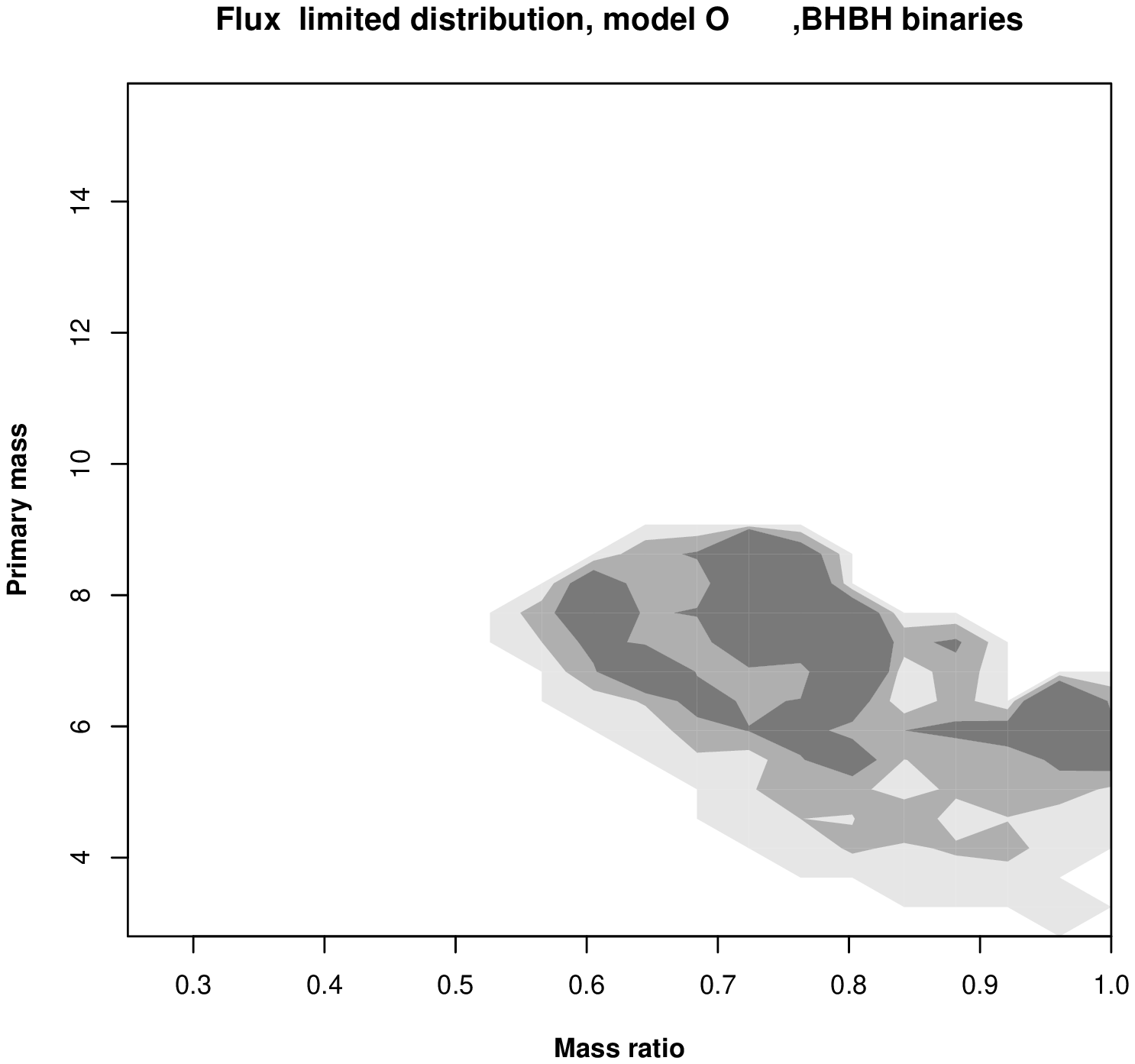}
\includegraphics[width=0.85\columnwidth]{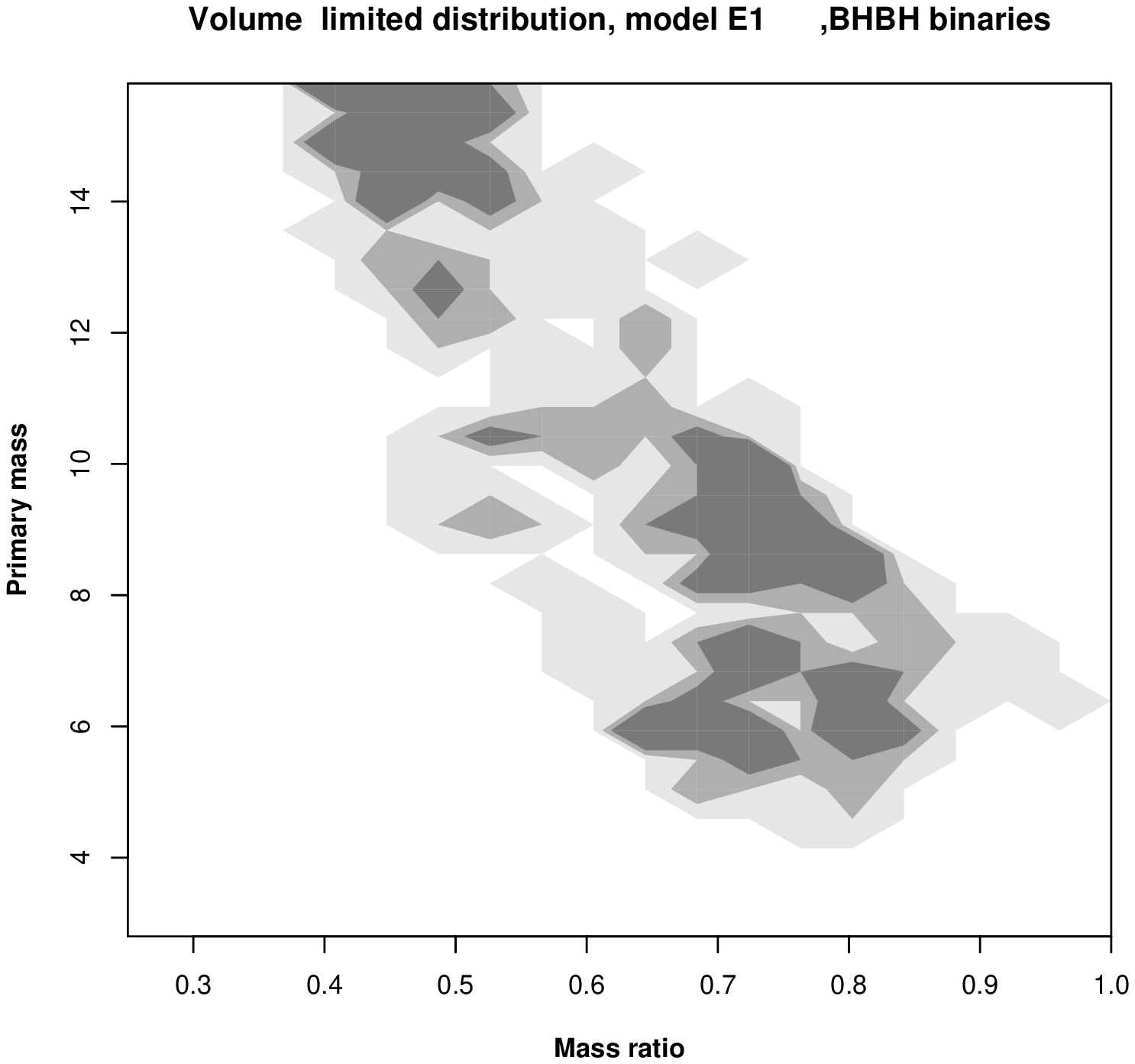}
\includegraphics[width=0.85\columnwidth]{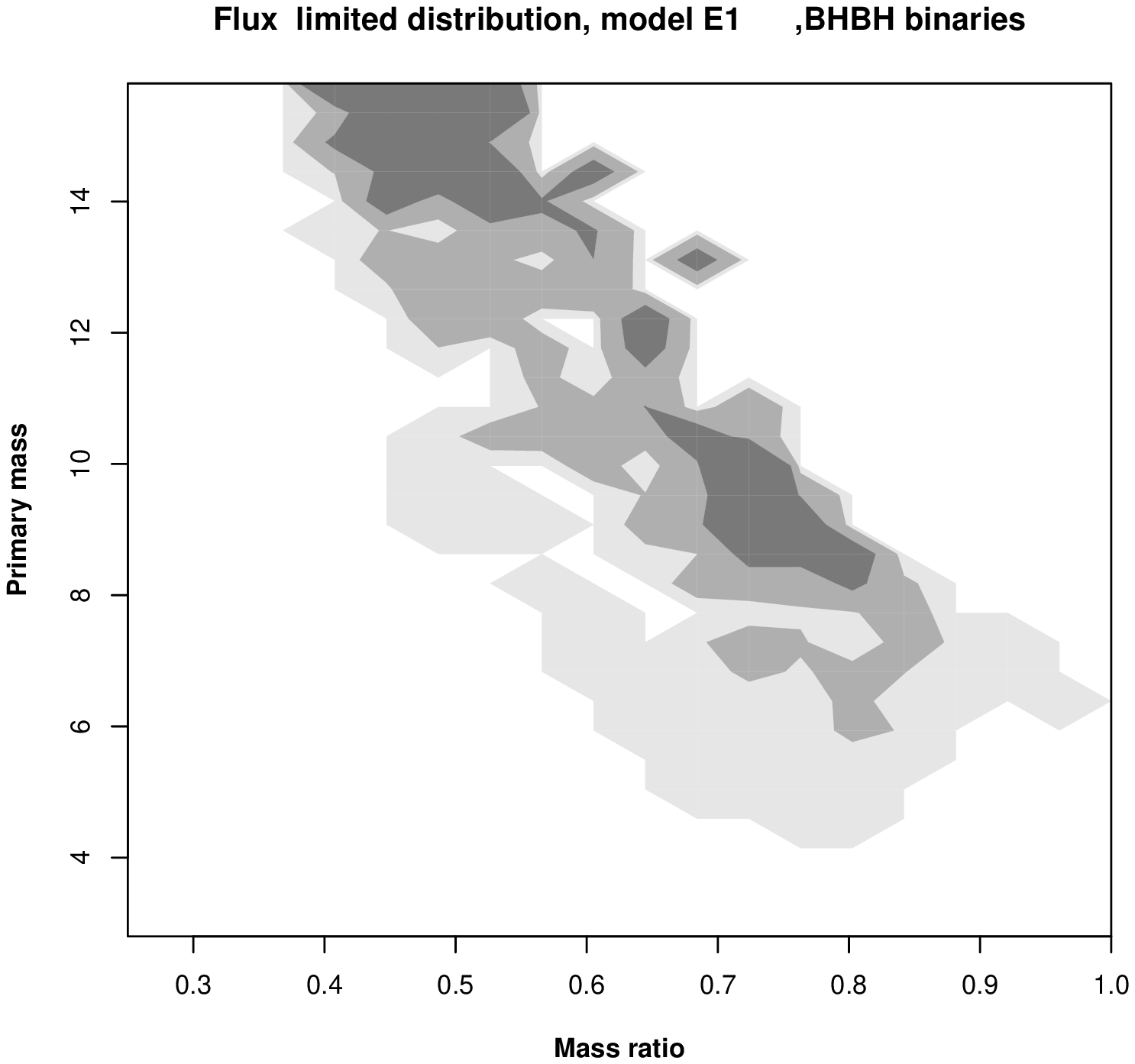}

\end{center}
\caption{The volume limited  and observability weighted distributions of 
of the parameters of BHBH binaries within model A (top panel),
and model O (bottom panel). The  region in dark gray encompasses
68\% of the systems, the medium gray corressponds to 95\%,
and the light gray corresponds to all binaries in the simulation. }
\label{bhbh}
\end{figure*}
\section{Results}

In the following we will assume that the  maximum 
mass of a neutron star is $3\,M_\odot$. All objects above 
this value will be considered as black holes. 
In our simulations the minimum mass of a 
neutron star is $1.2\,M_\odot$.
Thus we can classify all binaries as double neutron star (NSNS),
black hole neutron star (BHNS), or double black hole binaries (BHBH).
These three categories will be analyzed separately. 
We will present the distributions of binary parameters 
in the space spanned by the mass of the primary (the more
massive component of a compact object binary) and the mass ratio $q$.
 
\subsection{NSNS binaries}

We present the volume limited distributions of 
$q$ and $m_2$ in obtained in the framework of model A in the top panels
of Figure~\ref{nsns}.
The volume limited  distribution for the case of 
NSNS systems (top left panel) exhibits a peak   for the 
systems with nearly equal masses just above   the minimal mass 
of a neutron star.
 This roughly corresponds with the 
observations of pulsars where most systems have similar masses.
 There is however, a long tail in the distribution
extending to systems with large mass of the primary, and low mass ratio.
these are systems consisting of a neutron star with a mass near the 
maximum value and a companion neutron star with a  low mass.
Such systems have   chirp mass about $1.5$ times larger than the 
binaries from the above mentioned peak.
Therefore in the flux limited  distribution - top right panel in 
Figure~\ref{nsns} -these low mass ratio systems are showing up more 
prominently. It is however possible that the maximum mass of neutron star is
lower than $3\,M_\odot$ and some of the systems shown here 
harbor low mass black holes rather than neutron stars.  However even for 
the maximum mass of a neutron star of $2.0\,M_\odot$   there is 
still a large fraction of  low $q$ systems in the flux limited distribution.
We have examined all 21 model listed in Table 1 and nearly all the models show
a similar pattern in both the volume limited and flux limited distributions.
There is one exception - model C - for which we present the relevant distributions
in the bottom panels of Figure~\ref{nsns}. In this model we turn off the possibility
of hypercritical accretion onto a compact object in common envelope phase. 
This effectively shuts off the possibility of increasing significantly 
the mass of a neutron star through accretion. Therefore in the volume limited distribution 
there is quite a large concentration of systems with both masses below 
$1.5-1.6\,M_\odot$, and little number of binaries with low mass ratios.
Consequently in the flux limited distribution    the binaries with low 
mass ratio are nearly absent in contrast to model A. However, the
68\% contour includes systems with $q> 0.75$ and $m_2\approx 2.0\,M_\odot$, 
as well as some binaries with $q\approx 0.75$ and the mass of the
 primary close to the maximal mass of a neutron star in our model.

\subsection{BHNS binaries}

We present the volume limited and the flux limited distributions of $q$ and $m_2$ 
in Figure~\ref{bhns}.  The top panel of Figure ~\ref{bhns} corresponds to the standard model A.
The volume limited distribution shows a large number of binaries along a stripe 
stretching from $q\approx 0.7$ and $m_2\approx 4\, M_\odot$ to 
$q\approx 0.2$ and $m_2\approx 10\, M_\odot$. Systems above and to the right of this stripe would be 
classified as BHBH binaries. The volume limited distribution is dominated 
by binaries with low mass black holes. In the flux limited distribution 
the binaries with higher mass black hoes and low mass ratios start to play an important role.
This is due to  
the balance between the falling mass function and the increase in the 
chirp mass with increase of the mass of the primary. For nearly all models
 this leads to dominance of binaries where the black hole primary 
 has a mass between $6$ and $12\,M_\odot$, and the mass ratio is somewhere
 from $0.3$ for the most massive black holes to $0.5$ for the moderate mass
 ones. 
 
After examining the 21 models of Table 1  in the case of BHNS binaries one can 
distinguish two other classes of models with different distributions of binary parameters.
The first class consists of models , L2, M2, and O. For this models we show 
a representative case (model O) in the middle panel of Figure~\ref{bhns}. 
The common characteristic of this class of models is that the population of black holes in 
binaries lacks the very massive ones for various reasons. In the case of the model~O
shown in Figure~\ref{bhns} this is because we increase the range of masses 
of fall-back formation of black holes. Within the models E1, L2, M2 such binaries have a 
smaller chance of formation because of altering the treatment of the mass transfer events.
in consequence the is very little low mass ration BHNS binaries and  the
volume limited 
distribution is very similar to  the flux limited one.

 A separate class of consists of models E1, Z2 and G1. In their case high mass
 black holes are easily formed, because of decrease of strength of stellar winds (Z2 and G1).
 Model E1 favors survival of systems with high mass first born stars and leads
 to effective production of extreme mass ratio compact object binaries.  
 We present the two distributions for the case of model Z2 in the bottom panel
 of Figure~\ref{bhns}. Here the volume limited distribution is 
 dominated by a nearly vertical stripe at $q\approx0.2$. In the flux limited distribution
 the binaries with high mass primaries $m_2\approx 12\,M_\odot$ and 
 $0.1<q<0.3$ are dominant simply because of their high chirp mass.

 \subsection{BHBH binaries}

 The case of BHBH binaries is presented
 in   Figure~\ref{bhbh}. 
 The top panel corresponds to the  model A.
 The volume limited distribution 
 fills more or less uniformly the region allowed for the BHBH binaries. 
In the flux limited  distribution there is a preference 
for the high mass ratio (nearly equal mass) 
 and high mass ratio systems, i.e. these filling the top right corner of the
 plot. Thus the flux limited distribution is dominated by the
 systems with $q< 0.6$ and $m_2$ near the maximum mass
 produced in a given model. All models seem to follow this general trend.
 
 In  models O   the maximum mass of a black hole is
 decreased. We 
  present  the results of a calculation using model O, 
 in  the middle  panel of Figure~\ref{bhbh}.
 In these case the volume limited distribution also fills nearly uniformly 
 the region allowed for black holes. However because of it smaller size 
 the range of chirp masses for given mass ratio is not as large and 
 the flux limited distribution is only slightly shifted to higher masses 
 in respect to the volume limited one.    

 Another special case - model E1 - is presented in the bottom 
 panel of Figure ~\ref{bhbh}. Here because of lowered CE efficiency
 formation of equal mass compact object binaries is favored, while
 extreme mass binaries are preferentially formed. The volume limited distribution
 in this case is dominated by systems with mass ratio in the 
 range $0.4<q<0.6$ and a tail extending to $q=0.8$.

 Models of the class that favors production of massive
 black holes (Z2, G1) do not lead to qualitatively 
 different results than model A. In these models
 the flux limited distributions tend to concentrate around binaries with higher 
 mass ratio, $q> 0.7$ and higher total masses than in the standard model A.

\section{Conclusions}

We have calculated the expected distributions of masses
and mass ratios of compact object binaries to be observed
in gravitational waves. The results are based on the Star Track 
binary population synthesis code. 
For most of the models the observability weighted
distribution of   double neutron star systems
has two peaks: one with the mass ratio almost unity
and both masses near the smallest mass allowed for neutron stars, and another
with small mass ratio, consisting of stars with the mass near the maximum mass
of a neutron star in a binary with a star close to the minimum mass.
The reality of this second peak depends on the 
assumed maximum mass of a neutron star: the lower the maximum mass 
of a neutron star the smaller the small mass ratio peak.
The distribution of black hole neutron star binaries peaks at mass ratios
between $0.3$ and $0.5$.  The bulk of observed double black hole binaries
has mass ratios above $0.7$.  We have shown that these results
are rather generic and   depend weakly on the choice of a particular model 
of stellar evolution. 

The crucial parameter determining the shape of the distribution of 
the observed NSNS binaries is inclusion of the hypercritical accretion 
onto compact objects in common envelope events. In the case of BHNS and BHBH
binaries the most important parameters are these that alter the masses 
of the black holes in such binaries, and the common envelope efficiency. 
The masses may be altered  either due to the mechanism 
of compact object formation in supernova explosions, or due to 
particular treatment of mass transfer events. The distribution of 
the BHNS binary parameters is most sensitive to these changes.
However, we must note that hat the observed 
sample is dominated by the BHBH binaries. For most 
models more than 90\% of observed systems are double black 
hole binaries \citep{2003ApJ...589L..37B}.

 These results can be used as a 
guideline for choosing the initial conditions in numerical simulations
of mergers of compact object binaries. 
Additionally the results of this work can be used in 
preparing data analysis software using templates for detection of gravitational 
waves from compact object inspiral.  Coalescences of   BHBH binaries 
dominate the observed sample, and we find that the observability weighted 
distribution is peaked around nearly equal mass binaries.  
We find that in most  models the flux limited sample of NSNS
binaries contains a large fraction of non equal mass objects.
 We conclude 
that the initial search for gravitational waves from coalescences 
of compact object binaries should concentrate on BHBH coalescences
with mass ratio close to unity, and the low mass ratio NSNS coalescences
should be taken into account.

Finally, we   note that this work only includes binaries
that evolved in galaxies, and neglects all possible effects, like multiple
stellar interactions
that are relevant for evolution in dense stellar clusters.

\section*{Acknowledgments}
 
This work has been supported by the KBN grant 5P03D01120 
and Technology Programme EPAN-M.43/2013555 and the EU Programme
 ``Improving the Human Research Potential and the Socio-Economic
 Knowledge Base'' (Research Training Network Contract
 HPRN-CT-2000-00137).
We are grateful to the referee for the comments on this manuscript.

\newcommand{\mnras}{MNRAS}
\newcommand{\aap}{A\&A}
\newcommand{\aaps}{A\&A Supp.}
\newcommand{\apj}{ApJ}\newcommand{\apjl}{ApJ}
\newcommand{\araa}{ARAA}
\newcommand{\pasp}{PASP}
\newcommand{\prd}{Phys. Rev. D}
\newcommand{\nat}{Nature}

\bsp

\label{lastpage}

\end{document}